\documentstyle{article}
\include{epsf}

\begin{document}

\begin{center}

{\bf \large Template-stripped gold surfaces with 0.4 nm rms roughness 
suitable for force measurements.  Application to the 
Casimir force in the 20-100 nm range.}

\vspace*{2mm}

{Thomas Ederth\cite{byline}}

\vspace*{2mm}

{\em  Department of Chemistry, Surface 
Chemistry, Royal Institute of Technology, SE-100 44 Stockholm, 
Sweden\\
and Institute for Surface Chemistry, Box 5607, SE-114 86 Stockholm, 
Sweden}

\begin{abstract}
Using a template-stripping method, macroscopic gold surfaces with 
root-mean-square (rms) roughness $\leq$0.4 nm have been prepared, 
making them useful for studies of surface interactions in the 
nanometer range.  The utility of such substrates is demonstrated by 
measurements of the Casimir force at surface separations between 20 
and 100 nm, resulting in good agreement with theory.  The significance 
and quantification of this agreement is addressed, as well as some 
methodological aspects regarding the measurement of the Casimir force 
with high accuracy.
\end{abstract}

% \pacs{PACS numbers: 06.60.Ei, 12.20.Fv}

\end{center}

\section{Introduction}

More than 50 years ago, Casimir predicted that two parallel conducting 
plates attract each other in vacuum \cite{Casimir1948Attra}. 
The attraction is the result of a modification of the electromagnetic 
modes between the plates caused by the conducting boundaries.
The magnitude of this force per unit 
area between parallel plates, at separation $d$, is:
\begin{equation}
	{F\rm (\mit d\rm ) \over \mit A}\rm =-{{\mit \pi }^{\rm 2}\mit 
	\hbar c \over \rm 240{\mit d}^{\rm 4}}
\label{casimirFlats}
\end{equation}
Despite its implications in areas as diverse as cosmology, Rydberg 
atom spectroscopy, particle physics, and quantum field theory (see  
\cite{Mostepanenko1988Casim,Milonni1994TheQu} for reviews), 
quantitative experimental verification did not appear until very recently, when 
Lamoreaux investigated this force in the range 0.6-6 $\mu$m using a 
torsion pendulum \cite{Lamoreaux1997Demon}, and Mohideen and Roy used 
an atomic force microscope (AFM) for studies in the 0.1-0.6 $\mu$m 
regime \cite{Mohideen1998Preci,Roy1999Impro}.  The agreement with 
theory was claimed to be 5\% and 1\% in these experiments, respectively, but 
surface roughness, the use of multi-layer structures, and uncertainty 
regarding the absolute surface separation complicate the analysis in 
both cases. The experimental setups in these experiments were sphere-flat 
configurations (which is mathematically equivalent to the 
crossed-cylinder geometry used in this study).  For these cases, the 
proximity force theorem \cite{Blocki1977Proxi} (or ``Derjaguin 
approximation'' \cite{White1983OnThe}) can be 
used to transform the result for parallel flats, yielding instead for 
a sphere and a flat (or two crossed cylinders):
\begin{equation}
	{\mit F\rm (\mit D\rm )}\rm =-{{\mit \pi }^{\rm 3}\mit R \hbar c 
	\over \rm 360{\mit D}^{\rm 3}}
\label{casimirSphereFlat}
\end{equation}
\noindent where ${\mit D}$ is the closest separation between the 
bodies, and ${\mit R}$ is the radius of the sphere for the sphere-flat 
geometry, whereas for crossed cylinders $R\rm =\sqrt {{\mit R}_{\rm 
1}{\mit R}_{\rm 2}}$, where ${\mit R}_{\rm 1}$ and ${\mit R}_{\rm 2}$ 
are the radii of the cylinders.  The Casimir result holds for two 
smooth and perfectly conducting bodies interacting in vacuum at zero 
temperature, and considerable effort has 
been devoted to the derivation of corrections to 
Eq. (\ref{casimirSphereFlat}) for non-ideal experimental conditions.

The correction for finite temperature has different functional forms 
depending on the value of the parameter $t = \mit k_{\rm B}TD/ \hbar c $.  For 
the temperatures and separations considered here, $t$ 
\raisebox{-.5ex}{$ \stackrel{\scriptstyle <}{\scriptstyle \sim} $} 
0.01, which is in the low temperature regime 
\cite{Mehra1967Tempe,Schwinger1978Casim}. The relative magnitude of 
this correction is less than $10^{-4}$ in the range 20-100 
nm, and is apparently of little importance.

The deviations from the Casimir result due to finite conductivity have 
been estimated using a plasma model 
of the metal with dielectric function $\mit \varepsilon (\omega) \rm 
= 1 - \mit \omega_{p}^{\rm 2}/ \omega^{2}$, where $\omega_{p}$
is the bulk plasma frequency. The correction has the form of a series 
expansion in terms of $\lambda_{p}/D$
\cite{Schwinger1978Casim,Lamoreaux1997Demon,Bezerra1997Casim},
and has been determined at least to the fourth order \cite{Klimchitskaya1999Compl}.
At small separations, where the wavelengths of the lowest possible 
intersurface modes approach the plasma wavelength, the correction for 
finite conductivity based on the plasma model is no longer valid.  Lamoreaux 
\cite{Lamoreaux1999Calcu} calculated the interaction with Lifshitz 
teory \cite{Lifshitz1956Theor} instead, using spectroscopic data. This 
avoids using the plasma model conductivity correction, but instead 
introduces the difficulty of determining the frequency dependence of the
permittivity of the metal over a wide frequency range. In this 
report, where the separation range is $< \omega_{p}$, a similar 
method is used.

In the roughness correction by 
Klimchitskaya \cite{Klimchitskaya1996TheCor}, the corrugation amplitude 
$\mit {A}_{r}$ is chosen such 
that the deviation of the surface shape  
from the ideally smooth is described by $\mit z = A_{r}f_{r}(x,y)$, 
where $\rm max\left|{\mit f_{r}(x,y)} \right|=1$.
Thus, $\mit {A}_{r}$ should be taken as half the maximum 
peak-to-trough roughness over the surface, and assuming a random 
roughness distribution, the resulting 
correction is to second order:
\begin{equation}
	{{\mit F}^{R}\left({D}\right)}={F}\left({D}\right)
	\left({1 + 6{\left({{{\mit A}_{r} \over D}}\right)}^{\rm 2}}\right)
\label{roughCorr}
\end{equation}
Again, higher order contributions have been calculated for some 
experimental conditions \cite{Bordag1995TheCa}.

For further investigations of the Casimir force and 
related phenomena, improvements not only of the corrections, but also 
of the experimental procedures are 
required to obtain accurate results.  This report describes the 
implementation of a surface preparation procedure resulting in 
macroscopic metal surfaces with arbitrary thickness, whose roughness 
is about one order of magnitude smaller than reported in previous 
experiments.  The applicability of such surfaces to force measurements 
is demonstrated by measurements of the Casimir force at 
separations down to 20 nm.

\section{Numerical procedure}

\begin{figure}[tbp]
	\centering
	\epsfxsize=100mm \epsfbox{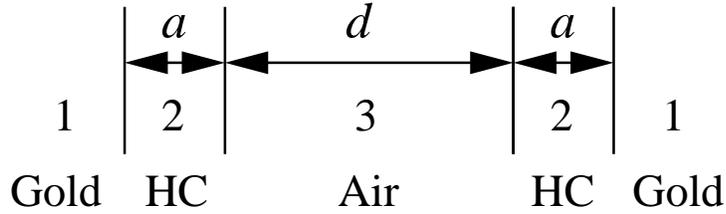}
	\caption{Schematic of the calculated system. The gold layers are 
	assumed semi-infinite, and the thickness $a$ of each hydrocarbon (HC) layer 
	is 2.1 nm. 
	The zero separation ($d$ = 0) refers to the point of contact of the 
	two hydrocarbon layers.}
	\label{fig:Layers}
\end{figure}

The interaction between the metal surfaces were calculated as follows:
For two gold plates with dielectric function $\varepsilon_{1}$, 
covered with hydrocarbon layers ($\varepsilon_{2}$, the purpose of 
these are explained further down) of thickness $a$, interacting across 
air (for which we assume $\varepsilon_{3} = 1$), the free energy of 
interaction per area at a separation $d$ is given by 
\cite{Mahanty1976Disper,Parsegian1973vande}:
\begin{equation}
	F\rm (\mit d\rm ,\mit T\rm )={{\mit k}_{B} \mit T \over \rm 8\mit 
	\pi {d}^{\rm 2}} \sum_{\mit n\rm =0}^{\infty } {'} \mit I\rm 
	({\mit \xi }_{\mit n}\rm ,\mit d\rm ) \rm , \hspace{5mm} \mit 
	\xi_{n} = {{\rm 2} \pi n k_{\rm B} T \over \hbar}
\label{Lifshitz}
\end{equation}
where the prime on the summation means that the term $n = 0$ should be 
halved. The separation $d$ is taken to be zero where the hydrocarbon layers 
contact each other, see Figure~\ref{fig:Layers}. Further,
\begin{eqnarray}
	\nonumber & I\rm ({\mit \xi }_{\mit n}\rm ,\mit d \rm ) = 
	{\left({{2{\mit \xi }_{\mit n}\mit d \over \mit c}}\right)}^{\rm 
	2} \int_{1}^{\infty } \rm \Bigl\{ \ln 
	\left({1-{\left({{\overline{\rm \Delta }}_{\rm 31}}\right)}^{2} 
	\exp {\left(-{2\mit p{\xi }_{n}d \over \mit c}\right)}}\right)\rm & \\
	\nonumber & \mbox{} \rm + \ln \left({1-{\left({{\rm \Delta }_{\rm 
	31}}\right)}^{2} \exp {\left(-{2\mit p{\xi }_{n}d \over \mit 
	c}\right)}}\right) \Bigr\} \mit p\rm d \mit p & \\
	 \label{LifInt}
\end{eqnarray}
where
$$ {\overline{\rm \Delta }}_{\rm 31}=
{{\overline{\rm \Delta }}_{\rm 32} + {\overline{\rm \Delta}}_{\rm 21}
\exp \left({ -{2 \mit {\xi }_{n} a s_{\rm 2} \over \mit c}}\right)
\over
1+ {\overline{\rm \Delta}}_{\rm 32} {\overline{\rm \Delta}}_{\rm 21}
\exp \left({ -{2 \mit {\xi }_{n} a s_{\rm 2} \over \mit c}}\right)} $$
(similarly for ${\rm \Delta }_{\rm 31}$). For any two adjacent layers $i$ and $j$
$${\overline{\Delta }}_{ij}={{s}_{j}{\varepsilon 
}_{i}- p{\varepsilon }_{j} \over {s}_{j} 
\varepsilon_{i}+ p \varepsilon_{j}} \rm , 
\hspace{2mm}
\mit {\rm \Delta }_{ij}={s_{j} - p \over s_{j} + p} 
\rm , \hspace{2mm}
\mit s_{j} = \sqrt{p^{\rm 2} - {\rm 1} + \varepsilon_{j}}$$
where $\varepsilon_{i} = \varepsilon_{i} \rm ( \mit i \xi \rm)$.
The dielectric function has a real and an imaginary component,
$\varepsilon \rm (\mit \omega \rm )=\mit \varepsilon \rm '(\mit 
\omega \rm 
)+\mit i\varepsilon \rm ''(\mit \omega \rm )$.
For a given frequency
$\rm \varepsilon '+\mit i\varepsilon \rm ''={\mit n}^{\rm 2}-{\mit 
k}^{\rm 2}+\mit i\rm 2\mit nk$,
but only the imaginary part of the dielectric function is required to 
calculate $\varepsilon \rm ( \mit i \xi \rm)$ along the imaginary 
axis, using  the Kramers-Kronig 
relationship:
\begin{equation}
	\varepsilon \rm (\mit i\xi \rm )=1 + {2 \over \mit \pi }\int_{\rm 
	0}^{\infty }{\mit x \varepsilon \rm ''(\mit x\rm ) \over {\mit 
	x}^{\rm 2}+{\mit \xi }^{\rm 2}}\rm d \mit x
	\label{KK}
\end{equation}
Tabulated spectroscopic data ($n$ and $k$) for gold \cite{Palik1985Hand} 
was used to calculate $\varepsilon \rm ( \mit i \xi \rm)$ using 
(\ref{KK}) for each frequency $\xi_{n}$.  In the 
low-frequency regime, the dielectric function was extrapolated using a 
Drude model:
\begin{equation}
	\varepsilon \rm (\mit i \xi \rm )=1+{{\mit \omega }_{\mit p}^{\rm 2} \over 
	\rm (\mit \xi^{\rm 2} \rm +\mit \gamma \xi \rm )}
	\label{eq:Drude}
\end{equation}
The plasma frequency $\omega_{p}$ = 1.4 $\times 10^{16}$ and 
the relaxation parameter $\gamma$ = 5.3 $\times 10^{13}$ were obtained as described in  
\cite{Lambrecht1999Casim}.
The optical properties of the hydrocarbon layer were modelled with a 
single oscillator \cite{Mahanty1976Disper}:
\begin{equation}
	\varepsilon \rm (\mit i\xi \rm )=1+{(\mit n^{\rm 2} \rm -1) \over 1+(\mit 
	\xi \rm /{\mit \omega }_{\rm UV})}
	\label{HC}
\end{equation}
where $n = 1.5$ and $\omega_{\rm UV} = 3.0 \times 10^{15}$ for 
a solid hydrocarbon \cite{Israelachvili1985Inter}.  Beyond 
the plasma frequency ($\omega_{p}^{\rm 2} = \mit Ne^{\rm 
2}/\varepsilon_{\rm 0}m_{e}$) the plasma model was used.
The total interaction does not depend critically on the 
hydrocarbon layer, and more elaborate models did not produce 
significantly different results.

For gold and hydrocarbon, $\varepsilon \rm ( \mit i \xi \rm)$ was 
calculated from $10^{14}$ to $10^{19}$ rad/s, by integration of 
(\ref{KK}) between $10^{12}$ to $10^{21}$ rad/s for each frequency 
$\xi$.  The integral (\ref{LifInt}) was then evaluated for $p$ between $1$ and 
$10^{4}$, and the summation in (\ref{Lifshitz}) continued until 
doubling the number of terms resulted in a change of less than 0.01\%.

To fit the calculated interaction to the measured data the function
\begin{equation}
	\rho = \rm { \left( \mit F_{exp}(d+\delta) - F_{calc}(d) - {\alpha \over d} 
	\rm \right) }^{2}
	\label{Fit}
\end{equation}
was minimized with respect to $\delta$ and $\alpha$.  The first term 
on the right hand side
is the measured force, where the parameter $\delta$ is the deformation 
of the surfaces along the symmetry axis, and is used to obtain the 
true surface separation.  The second term is the calculated 
interaction as described above, corrected for surface roughness 
to the second order.  The last term is the electrostatic 
force between the surfaces caused by residual potential differences.

\section{Experimental procedure}

The gold surfaces were prepared by a template-stripping method adapted 
from Wagner \cite{Wagner1995Forma}.  Thin (10-15 $\mu$m) 
freshly cleaved mica sheets were cut in $\rm 1 \times 1 \ {cm}^{2}$ 
pieces using a hot platinum wire, and a 200 nm gold layer was 
deposited onto the mica in an ultra-high vacuum evaporator at a rate 
of 0.5 nm/s, with the evaporation pressure typically at $3 \times {10}^{-8}$ Torr 
(considerably thicker gold layers can be prepared in the same manner 
with no differences in subsequent preparation steps, the roughness of 
the final gold surface remains the same).  The gold-coated 
mica pieces were glued (Epo-Tek 301-2, Epoxy Technology) gold-side 
down onto cylindrical silica discs (R = 10 mm).  The day before use, 
the discs were immersed in tetrahydrofurane until the mica sheet came 
loose (a few minutes).  After drying in a gentle N$_{2}$ flow, 50 
$\mu$m gold wires were attached to the bare gold using a gold spring 
clip, whereupon the surfaces were immersed into a 1 mM solution of 
hexadecanethiol (Fluka, 95\%) in ethanol, and
incubated overnight.  The hexadecanethiol self-assembles into a 
close-packed crystalline monolayer, with the hydrocarbon chains facing 
outwards and the thiol covalently attached to the gold substrate 
\cite{Dubois1992Synth}.  This layer prevents contaminants from the 
laboratory atmosphere to adsorb onto the surface 
\cite{Smith1980TheHy}, and so serves to keep the surface well-defined, 
which is necessary for estimating surface deformation in the force 
measurements.  It also prevents cold welding of clean gold layers in 
contact, which would damage the surfaces upon separation.  The 
thickness of each thiolate layer is approximately 2.1 nm 
\cite{Atre1995Chain}.  After removal from the thiol solution, the 
samples were sonicated in ethanol to remove physisorbed thiols.  The 
surfaces were then mounted in a crossed-cylinder configuration in the 
force measuring device, and the wires from the two surfaces were 
connected with a gold clip, providing an all-gold conducting path 
between the surfaces (in such a way that the movement of one surface 
is not transmitted to the other surface through the wire).

The surface roughness was measured with an AFM (Nanoscope III, Digital 
Instruments) in tapping mode.  The roughness parameters are as 
measured over $\rm 1 \times 1 \ \mu {m}^{2}$, and evaluated using the 
software supplied with the instrument.

The contact angles with water were determined by slowly expanding a droplet 
on a flat template-stripped hydrocarbon covered surface, and 
determining the angle formed between the water droplet and the 
substrate with a microscope goniometer (Rame-Hart NRL 100).

\begin{figure}[tbp]
	\centering
	\epsfxsize=100mm \epsfbox{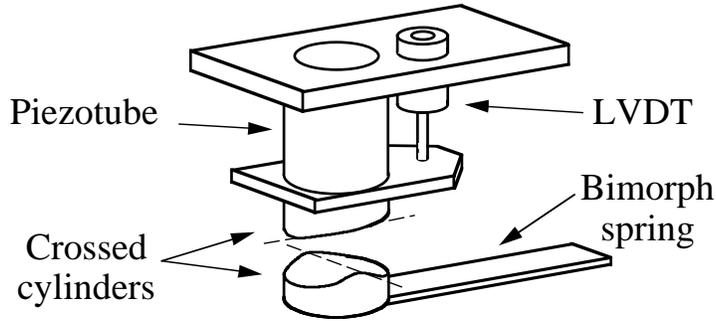}
	\caption{Simplified view of the force measuring device. The position of the 
	upper surface is controlled with a motorized stage (not in figure) and a 
	piezoelectric tube, while the response of the lower is detected with 
	the bimorph transducer, acting as the measuring spring. The LVDT is 
	used to monitor the non-linear expansion of the piezotube. The radius 
	of curvature of the cylindrical surfaces is 10 mm.}
	\label{fig:Masif}
\end{figure}

The force measurement device (Figure~\ref{fig:Masif}) works in a 
manner similar to the AFM, but is designed for measurements between 
macroscopic surfaces \cite{Parker1994Surfa}.  One surface is mounted 
onto a piezoelectric tube, whose position can be adjusted with a 
motorised translation stage to within $\pm$ 50 nm.  A linearly 
variable displacement transducer (LVDT) is mounted in parallel with 
the piezo to measure the tube expansion, in order to eliminate 
piezotube hysteresis in the subsequent data analysis.  The other 
surface is mounted onto a piezoelectric bimorph deflection sensor 
\cite{Stewart1995TheUse}, acting as a single cantilever spring, and the 
charge produced by the bimorph upon deflection is detected with an 
electrometer amplifier.  A force-distance profile is acquired by 
moving the surfaces towards each other at a constant rate from a 
separation $<$ 3 $\mu$m, using the piezotube.  When the surfaces 
contact each other, the surfaces are moved a further 200-300 nm 
together while being in contact (and the expansion of the piezotube is 
directly transmitted to the bimorph), before they are separated again.  
The average approach rate was approximately 80 nm/s.  The distance 
resolution was $\sim 0.1$ nm, and the force resolution $\sim$ 10 nN. 
Data is presented as equivalent free energy of interaction, 
i.e.\ force normalised with 2$\pi\times$radius (F/$2\pi$R); the 
normalised force resolution is $\sim$ 0.1 $\mu$N/m (or $\mu$J/m$^{2}$).
The force profiles were averaged by arranging the force-distance data 
pairs from five individual approaches into a single column, sorting 
the data by distance order and calculating a running 
average.  All experiments were performed in air at 25 $\pm$ 1 
$^{\circ}$C, and the relative humidity during the experiments 
was $\leq$ 60\%. A set of external caliper gauges with a precision of $\pm 
0.05$ mm were used to determine the radii of the surfaces after the 
experiments.  The relatively stiff mica templates used to fix the low 
viscosity glue in the preparation step, ensure that the deviations of 
the local radii from the macroscopic radii are small.  The studied 
separation range was determined by the force measurement device: the 
force resolution limit approaches the magnitude of the calculated result 
at separations beyond 100 nm, and at about 
20 nm, the gradient of the force is comparable to the stiffness of the 
measuring spring, and the surfaces ``jump'' into contact.

\section{Results}

\subsection*{Surface preparation}

\begin{figure}[tbp]
	\centering
	\epsfxsize=100mm \epsfbox{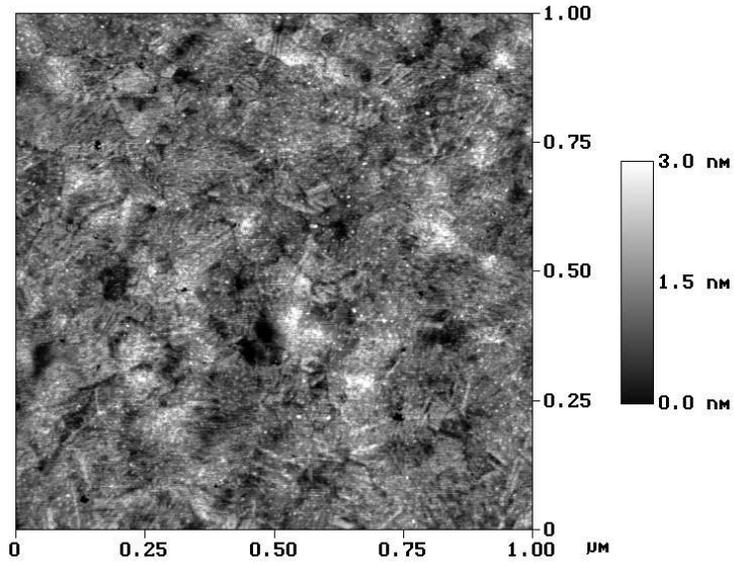}
	\caption{Atomic force microscope image showing the structure of 
	the template-stripped gold surfaces.  The peak-to-trough roughness 
	over $\rm 1 \times 1 \ \mu {m}^{2}$ areas are 3-4 nm, the 
	corresponding rms roughnesses 0.3-0.4 nm.}
	\label{fig:Tsg}
\end{figure}

AFM investigation of the template-stripped gold surfaces reveal 
peak-to-trough roughness of 3-4 nm, with corresponding 
root-mean-square (rms) roughness in the 0.3-0.4 nm range (see 
Figure~\ref{fig:Tsg}).  A significant contribution to the 
peak-to-trough value comes from a sparse population of pinholes in the 
layer, probably resulting from insufficient annealing or heterogeneous 
growth of the gold layer during the initial stages of the evaporation.  
Comparing the results with those of Wagner \cite{Wagner1995Forma}, it 
appears that annealing the films after evaporation might yield a 
further reduction of the roughness.  Compared to the 3 nm rms 
roughness amplitude reported by Roy in a recent report using a smooth 
metal coating \cite{Roy1999Impro}, the template-stripping method 
yields a reduction of the roughness with almost one order of 
magnitude.  With this roughness amplitude, however, the second order 
roughness correction is still $\sim$ 20\% at 20
nm, and the calculated Lifshitz result must be corrected accordingly.

The contact angles with water after adsorption of the thiolates was 
110$\pm$2$^{\circ}$, indicating that the surfaces expose a dense 
hydrocarbon layer.

\subsection*{Force measurements}

\begin{figure}[tbp]
	\centering
	\epsfxsize=100mm \epsfbox{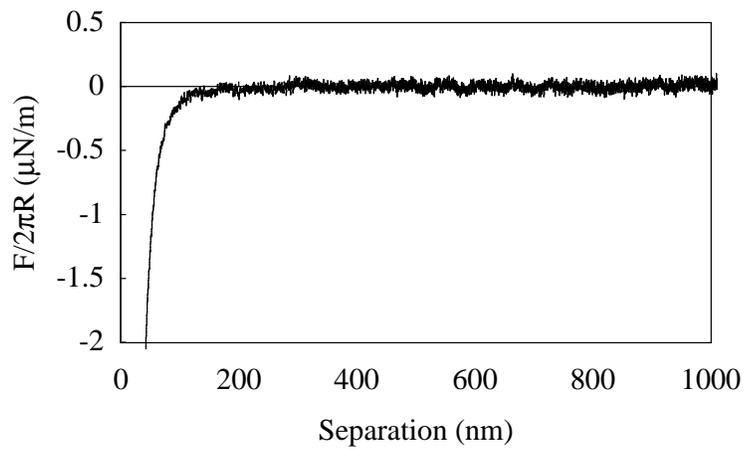}
 	\caption{A Force-distance profile for a single approach, the 
	displayed interval comprises approximately 7000 data points.}
	\label{fig:Long}
\end{figure}

A force-distance profile for a single approach is shown in 
Figure~\ref{fig:Long}. There 
appears to be no significant electrostatic interaction at large 
separations, which is also confirmed by the result of the fitting 
procedure (see further down).  However, the used method provides only 
indirect determination of the separation between the surfaces, and the 
distance scale has to be corrected for deformation of the surfaces 
caused by attractive forces when they are in contact.  The relatively 
soft glue used to support the gold layer causes the surfaces to deform 
substantially, but the layered structure of the surface makes direct 
application of continuum theories for surface deformation questionable 
\cite{Johnson1971Surfa}.  The central displacement $\delta$, i.e.\ the total 
compression of the two surfaces along the symmetry axis 
has been calculated using finite element analysis 
for the silica-glue-gold system under consideration, and was found to 
be 18-20 nm for surfaces with the glue thicknesses used here 
\cite{Sridhar1997Adhes}.  This implies that the measured force 
profiles should be shifted 18-20 nm towards shorter separations.

Figure~\ref{fig:Casimir} shows two force profiles calculated as 
averages of five different approaches in two independent experiments.  
Fitting the averaged data to the Lifshitz result using (\ref{Fit}) 
yields a total compression $\delta$ of 9 and 12 nm, respectively, in 
moderate agreement with the numerical result.  However, the calculated 
value of the compression corresponds to the contact of ideally smooth 
surfaces, while the finite roughness of the real surfaces reduces the 
adhesion (and the central displacement), and the calculated value must 
be used as an upper bound to the actual central displacement.  
Taking this into consideration, the deviations are perfectly 
reasonable.

The parameter $\alpha$ measuring the electrostatic contribution to the 
force is $<1.3 \times 10^{-23}$ Nm for both data sets, which results in 
an electrostatic force of the order of the instrument resolution at 
the shortest separation, and it is concluded that this contribution to 
the total interaction can be ignored (replacing the 1/$d$ term with a 
1/$d^{\rm 2}$ term, taking patch charges into account, does not 
improve the fit).

The absence of charges on the dielectric hydrocarbon surface might be 
surprising, but is probably a result of the natural humidity in the 
air surrounding the surfaces.  At the relative humidities (RH) where 
the experiments were performed ($\leq$ 60\%), the amount of water 
adsorbed from the atmosphere onto
the non-polar hydrocarbon layers is small, however. For similar surfaces the 
water coverage at 100\% RH has been determined to 0.8 monolayers
\cite{Thomas1999Water}.  For solid polyethylene with higher 
affinity to water (contact angle $\theta$ = 88$^{\circ}$), water layers of the order 
of 0.1 nm at 60\% RH have been reported \cite{Tadros1974Adsor}, while
a surface conductivity study arrived at a 3 monolayer water thickness 
at 100\% RH for a surface with $\theta$ = 104$^{\circ}$ 
\cite{Awakuni1972Water}.  Thus, assuming a 0.1 nm thick water layer on 
the surfaces appears to be a pessimistic estimate, and the effect on 
the interaction of such a layer was calculated using an oscillator 
model for water, where a Debye relaxation term in the microwave region 
is added to damped harmonic oscillators in the IR (5 terms) and UV (6 
terms) regions, with parameters as described by Parsegian 
\cite{Parsegian1975LongR} and Roth \cite{Roth1996Impro}:
\begin{equation}
	\varepsilon \rm (\mit i\xi \rm )=1 + \mit 
	{f \over \rm 1+\mit g \xi}\rm 
	+ \mit \sum\nolimits\limits_{j}^{} {{f}_{j} \over {\omega 
	}_{j}^{\rm 2}+{\xi}^{\rm 2}+{g}_{j}\xi }
	\label{Water}
\end{equation}
For equivalent separations between the solid surfaces, the effect of 
such a water layer corresponds to an increase in the calculated 
interaction of $\approx$ 1\% at 20 nm.  Considering the assumption of
a rather thick
water layer, the error introduced by neglecting this in the 
calculations is therefore concluded to be small.

After fitting the data, the total rms force deviation is $<$ 1\% of 
the force at 20 nm. The small differences between the two experimental 
data sets in Figure~\ref{fig:Casimir} after fitting indicate that the 
precision (repeatability) in the measurements is good, in fact 
as good as the agreement with the calculated interaction (the 
accuracy), using the same measure as above.

\section{Discussion}

\subsection*{The precision of the measurement}

Although the rms deviation between the experimental and 
theoretical results is $<$1\% at the shortest separation, 
it appears that this result cannot -- for several reasons -- be taken 
as confirmation of the theory at the same level of agreement.  First, 
the Lifshitz calculations based on optical data is insecure in that 
the optical data is incomplete, and extrapolations have to be made; 
the potential errors due to the choice of optical models in the 
extrapolated regimes (and the parameters used to describe them) have 
been reported recently \cite{Lambrecht1999Casim,Bostrom2000Comme}, and 
to ensure that correct data is used, spectroscopic 
data should be collected for the very surfaces that are used in the 
force experiments.

Further, from the series 
of reports by Mohideen and co-workers 
\cite{Mohideen1998Preci,Mohideen1999Reply,Klimchitskaya1999Compl}, it 
seems that the relative rms error at the shortest separation is
too blunt a measure of the agreement between theory and experiment: 
the first analysis of their experiment in the range 100-900 nm used a 
method where the Casimir force corrections to second order for 
conductivity, roughness and for the finite temperature were multiplied 
together, resulting in an rms deviation (as calculated over the whole 
interaction range) corresponding to 1\% of the force at the shortest 
separation \cite{Mohideen1998Preci}.  This analysis was criticized by 
Lamoreaux \cite{Lamoreaux1999Comme}, claiming that the agreement must 
be coincidental, since the corrections for conductivity and roughness 
were not sufficiently detailed, and that the potential error caused by 
this might be greater than 50\%.  Subsequently, a different
theory, including the roughness and conductivity corrections to fourth 
order (and some ``cross-terms'' as well), and using a more elaborate 
quantitative description of the surface roughness, was used to produce 
a similar 1\% relative precision at the shortest
separation for the same data \cite{Klimchitskaya1999Compl}.  Thus, 
since it is emphasized in 
\cite{Mohideen1998Preci,Mohideen1999Reply,Klimchitskaya1999Compl} that 
no adjustable parameters were used, it seems that a 1\% rms agreement 
at the shortest separation allows for
erroneous models to fit the data, and should perhaps be considered an 
inappropriate criterion for agreement between theory and experiment.

\begin{figure}[tbp]
	\centering
	\epsfxsize=100mm \epsfbox{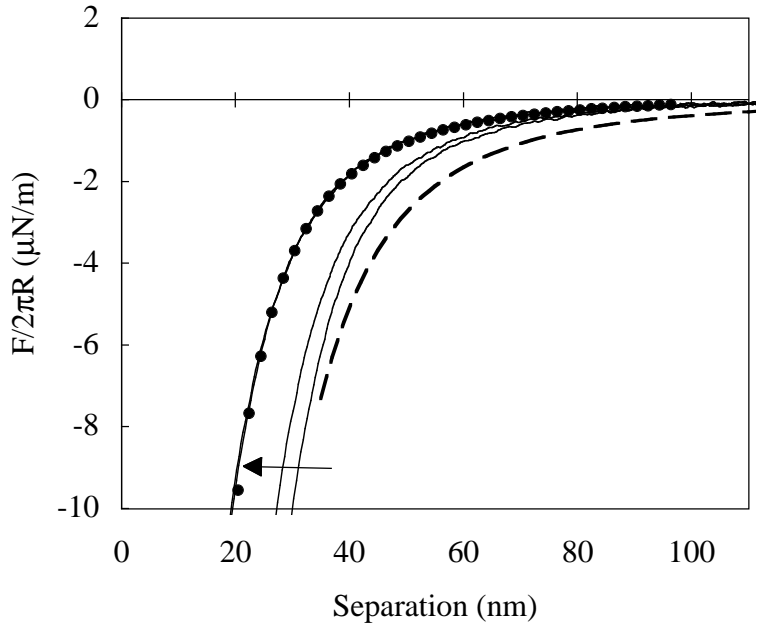}
	\caption{The solid curves under the arrow represent two 
	independent experiments: each of them is an average of 5 
	approaches.  To compensate for surface deformation, they are 
	shifted towards shorter separations when fitted with the 
	calculated interaction for the gold-hydrocarbon-air system 
	($\bullet$), where they coincide.  The deformations are 9 and 12 
	nm, respectively, in fair agreement with calculations (see text 
	for details).  The dashed line is the Casimir result, Eq.  
	(\ref{casimirSphereFlat}).}
	\label{fig:Casimir}
\end{figure}

\begin{figure}[tbp]
	\centering
	\epsfxsize=100mm \epsfbox{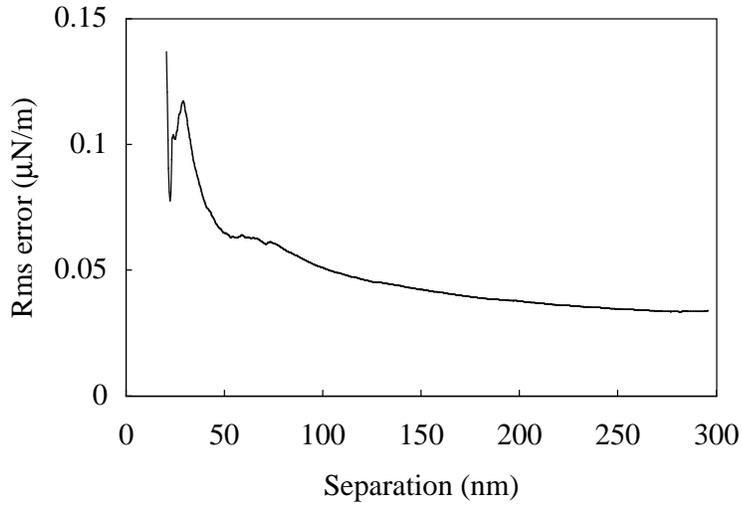}
	\caption{The accumulated rms error for one of the data sets in Figure~\ref{fig:Casimir}. 
	The error is calculated over ranges from 20 nm to the separations 
	indicated on the abscissa. The rms error relative to the 
	magnitude of the force at the shortest separation (20 nm) is 1\% for 
	the rms error calculated between 10 and 33 nm, decreasing to 0.36\% if 
	averaging is continued to 300 nm, even though 
	the relative measurement error increases steadily in this range, see 
	Figure~\ref{fig:RelErr}.}
	\label{fig:Error}
\end{figure}

\begin{figure}[tbp]
	\centering
	\epsfxsize=100mm \epsfbox{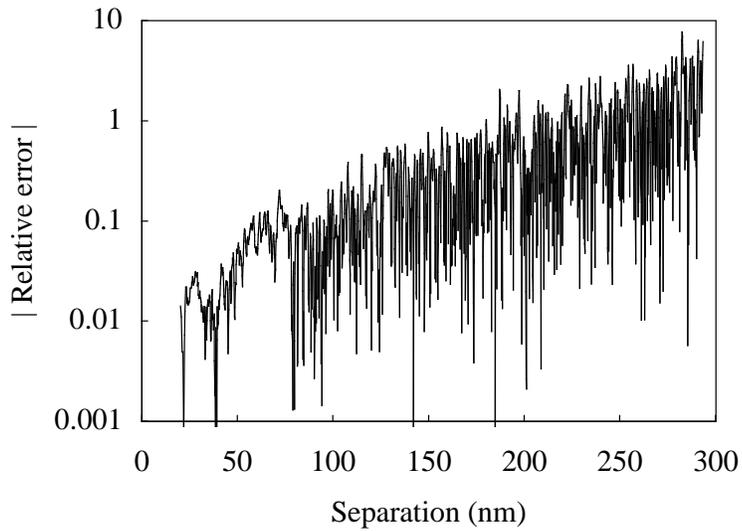}
	\caption{The magnitude of the relative error for the data set used to 
	calculate the rms error in 
	Figure~\ref{fig:Error}.}
	\label{fig:RelErr}
\end{figure}

One reason for this is that the rms error calculated as 
$\sigma = \sqrt{\sum(F_{exp}-F_{calc})^{\rm 2}/N}$ is unsuitable for relative error estimates for 
non-linear functions with wide variations in magnitude; 
even though the relative error in the measurement can amount to 100\% or more at large 
separations where the magnitude of the measured force approaches the 
resolution of the instrument, the average of those will be a small 
absolute error when measured relative to the magnitude of the force at 
small separations.  Averaging further into the region of low magnitude 
will continually decrease the rms error. For the data presented in 
Figure~\ref{fig:Casimir}, the rms error decreases if the spearation 
range used for the calculation is increased, as is clear from 
Figure~\ref{fig:Error}.  The rms deviation relative to the force at 
the shortest separation is 0.48\% if the deviation is computed from 20 
to 100 nm, decreasing to 0.36\% if the rms error summation is 
continued to 300 nm instead; indeed a meaningless measure of the 
accuracy.  At approximately 100 nm, the calculated force is of the 
order of the resolution of the instrument, and beyond this point the accumulated rms 
error decreases monotonically, even though the relative error in the 
measurement is steadily increasing, see Figure~\ref{fig:RelErr}.
For a single 
figure to measure the deviation between theory and experiment, the 
error at a particular separation is probably better weighted with the 
magnitude of the force, and the averaging 
certainly should not continue beyond the point where the magnitude of 
the calculated interaction approaches the noise level. The 
rms figures provided in \cite{Klimchitskaya1999Compl} show a similar 
trend; for deviations 
measured over 30, 100 and 223 data points (corresponding to 
separations 80-200, 80-460 and 80-910 nm) the rms error is 1.6, 1.5 
and 1.4 pN, respectively.

Whenever the separation between the surfaces is not measured directly 
(and with high accuracy), the uncertainty in the location of the 
measured curve along the separation scale will always be a source of 
error.  In the experiments presented here, the deformation of the 
surfaces is the only remaining fit parameter of significance, but it is not possible 
to establish with certainty that the central displacement $\delta$ 
obtained through the fit procedure is correct, which diminishes the strength of the 
measurement as a test of the Casimir force, and also precludes a 
quantitative assessment of the agreement between theory and 
experiment.

To determine the merit of the corrections to the Casimir force, a 
precise determination of the separation is essential; the corrections 
for conductivity and roughness are both expansions in 1/$d$, 
increasing their effect at shorter separations.  If there is 
uncertainty in the separation, the error caused by using the wrong 
theory is easily obscured by a shift along the separation axis, which 
corrects for the deviations at small separations where the errors are 
greatest, while making little difference at larger separations where 
the force profile is much flatter.  If, using the data in 
Figure~\ref{fig:Casimir}, the roughness correction is ignored in the 
calculated force profile, the total rms deviation at the smallest 
separation is 0.49\% for the rms calculated in the
range 20-100 nm, provided the experimental data is shifted 3.1 nm 
along the separation axis.  Without shifting the curve, the rms 
error can be kept $<$1\% if averaging is continued to 400 nm.
Besides, the data in 
Figure~\ref{fig:Casimir} can be shifted 0.5 nm in either direction, 
still keeping the rms error $<$1\% for averages between 20 and 100 nm.

The ambiguity due to the fact that the rms error is continuously 
decreasing as it is calculated over larger separations, implies
that the relative rms error at the 
shortest separation is an unsuitable measure of the agreement between 
theory and experiment, and the 1\% level, in particular, is too broad 
to discriminate the second order roughness correction from no 
correction at all, even though the magnitude of the forces differ 
with as much as 20\%.

In a similar fashion, the Au/Pd layers covering the Al surfaces used 
in \cite{Mohideen1998Preci,Roy1999Impro,Klimchitskaya1999Compl} were 
ignored in the analysis, but could be accomodated as additional layers 
with effective permittivity $\varepsilon \approx$ 1.2 without changing 
the rms error, provided all separations are increased 3 nm 
\cite{Klimchitskaya1999Compl}.  Incidentally, the plasma wavelength 
used in 
\cite{Mohideen1998Preci,Mohideen1999Reply,Klimchitskaya1999Compl}, 
$\lambda_{p} =$ 100 nm, was taken from \cite{Palik1985Hand}, which 
gives ``$\sim 15$ eV'' as the plasmon energy, corresponding to 
$\lambda_{p} \approx$ 83 nm.  Already this difference causes a $>$3\%
deviation in the conductivity correction used in 
\cite{Mohideen1999Reply}, where at the same time it is mentioned that 
``Small changes in $\lambda_{p}$ will not significantly modify 
$\sigma$.'', which appears to confirm that $\sigma$ is not a 
very good measure of the accuracy.

\subsection*{The methodological improvement}

The principal methodological improvement in this report is the 
preparation of metal surfaces with reduced surface roughness, though 
other problems common to this and the experiments discussed hitherto 
remain: the unknown absolute surface 
separation, the effect of additional layers, the determination of the 
permittivity (or finite conductivity correction) of the metals, and the 
presence of other interactions (principally electrostatic 
contributions).  The suggested 
procedure does not avoid these problems, but the two first points 
deserve some attention.

The use of macroscopic surfaces improves accuracy, since the magnitude 
of the involved forces are greater, but instead entails enhanced 
surface deformation problems.  Any two bodies in contact deform to an 
extent determined by a balance between the reduction in surface energy 
and the elastic strain energy caused by the deformation; for the 
gold-glue-silica system the deformation at surface contact (with zero 
applied external load) was calculated to be 18-20 nm.  If the two 
crossed cylinders used in the experiments were solid gold (all other 
things being equal), the calculated deformation would have been 7 nm 
(see the Appendix for details).  Klimchitskaya {\em et al.}\ mention 
that smoother metal coatings and surfaces with larger radii can be 
used to improve the precision of the measurements 
\cite{Klimchitskaya1999Compl}, but this will inevitably lead to 
increased problems with surface deformation.  It is a mistake to 
assume that this is a problem limited to the use of macroscopic 
surfaces only, but it ought to be a matter of concern also in the 
analysis of past AFM experiments 
\cite{Mohideen1998Preci,Roy1999Impro,Klimchitskaya1999Compl}.  If the 
surfaces are smooth and the interfacial energy of the contact is that 
of two hydrocarbon surfaces (which is about as low as is practically 
achievable in air or vacuum), a 200 $\mu$m polystyrene sphere (used in 
\cite{Mohideen1998Preci,Roy1999Impro}) interacting with a silica plate 
is compressed $\sim$10 nm upon contact, under zero applied load (see 
the Appendix for details).  Now, roughness decreases this figure since 
the effective contact area decreases, but on the other hand, the 
interfacial energy of a clean metal-metal contact might be two orders 
of magnitude higher than that for two hydrocarbon surfaces.  These 
issues have to be addressed if a proper estimate of the separation 
uncertainty is to be established.

The hydrocarbon layers described in this report were used both to keep 
the surfaces well-defined -- which is essential for deformation 
estimation -- and to avoid cold welding of the gold layers in contact.  
Thus, besides producing surfaces with low roughness, the proposed 
preparation procedure has the added advantages that the surface energy 
is well defined (and small) by use of the hydrocarbon layer, and that 
the theoretical treatment of this layer is fairly straightforward.

\section{Conclusion}

In conclusion, a template-stripping method was used to prepare smooth 
gold surfaces, with $\leq$0.4 nm rms roughness.  The roughness is 
independent of the thickness of the gold layer, and is about one order 
of magnitude smaller than surfaces used in previous experiments.  
These surfaces (covered with a hexadecanethiolate overlayer) were used 
to measure the Casimir force in air at separations between 20 and 100 
nm, a range that has previously been inaccessible due to the roughness 
of the samples.  The results were found to be in good agreement with 
the Lifshitz prediction for the interaction, once the deformability of 
the surfaces had been taken into account.  The experimental 
uncertainties, above all the deformation, makes a quantitative 
assessment of this agreement difficult, however.  Using the obtained 
data, it is also demonstrated that the rms error is a very ambiguous 
quantitative measure of the agreement between theory and experiment, 
and in particular that a 1\% level is not
cogent enough to discriminate the effect of corrections to the Casimir 
force.

\section*{acknowledgment}

The author thanks J. Daicic for discussions and valuable suggestions, 
B. Liedberg, Link\"oping University, for generous permission to use 
the surface preparation facilities in his laboratory, B.~W. Ninham for 
discussions, K.~L. Johnson and I. Sridhar for the deformation data, and the 
Swedish Natural Science Research Council for financial support.

\appendix
\section*{Surface deformation}

To calculate the deformation of elastic bodies in contact the models by 
Johnson {\em et al.} \cite{Johnson1971Surfa} (JKR) and Derjaguin {\em 
et al.} \cite{Derjaguin1975Effec} (DMT) are the most commonly used.  To discriminate the 
range of applicability of either model, a dimensionless parameter, 
$\mu$, is used \cite{Greenwood1997Adhes}:
\begin{equation}
	\mu \rm ={\left({{\mit R{\gamma }^{\rm 2} \over {\mit K}^{\rm 
	2}{\mit D}_{e}^{\rm 3}}}\right)}^{{1 / 3}}
\label{eq:Tabor}
\end{equation}
where $R$ is the radius of interaction as described in the 
Introduction, $\gamma$ is the interfacial energy of the contact, $K = 
[(1-\nu_{\rm 1}^{\rm 2})/E_{\rm 1} + (1-\nu_{\rm 2}^{\rm 2})/E_{\rm 
2}]^{\rm -1}$ contains two materials constants, the Young's modulus 
$E$, and the Poisson ratio $\nu$, for each material.  $D_{e}$ is the 
equilibrium separation between the surfaces in contact, which is 
difficult to establish, but a few \AA\ is a typical estimate.  For 
$\mu < 0.1$, that is for small and/or hard particles, the DMT model is 
appropriate, while the JKR applies where $\mu > 5$, and the 
interacting bodies are large and/or soft \cite{Johnson1997AnAdh}.  
Most macroscopic surfaces fall into the latter category, and so does 
the polystyrene spheres used in some recent AFM experiments 
\cite{Mohideen1998Preci,Roy1999Impro}.  For polystyrene spheres with
$R =$ 200 $\mu$m, Young's modulus $3\times10^{9}$ Pa and 
Poisson's ratio 0.33, interacting with a silica plate with $E = 8 
\times 10^{10}$ Pa and $\nu = 0.42$, and further assuming an equilibrium surface 
separation of a few, say, 3 \AA, and the interfacial energy of a
hydrocarbon-hydrocarbon contact, 0.05 J/m$^{\rm 2}$, which is
as low as is realistically obtainable, the parameter $\mu
\approx$ 12, which is in the JKR regime.

For the present purposes, the JKR result of most 
interest is the central displacement, $\delta$, i.e. the deformation along the 
symmetry axis under the externally applied load $F$ (where $F>0$ for 
compression):
\begin{equation}
	\delta \rm ={{\mit a}^{\rm 2} \over \mit R}\rm -{\left({  2\mit \pi 
	\gamma a \over K} \rm \right)}^{\rm 1/2}
\label{eq:JKR_cd}
\end{equation}
where $a$ is the radius of the contact region, given by
\begin{equation}
	{a}^{\rm 3}={\mit R \over \mit K}\rm 
	\left({\mit F\rm +3\mit \pi \gamma R\rm +{\left[{6\mit \pi \gamma 
	RF\rm +{\left({3\mit \pi \gamma R}\right)}^{\rm 
	2}}\right]}^{1/2}}\right)
\label{eq:JKR_cr}
\end{equation}
from which it is clear that the surfaces deform even without externally 
applied load. The pull-off force, the negative load that has to 
be applied to separate the surfaces from adhesive contact is:
\begin{equation}
	F_{a} = -{\rm 3 \over 2}\pi \gamma R
	\label{PullOff}
\end{equation}
which can be used to determine the interfacial energy of two 
interacting bodies.

% \bibliography{Casimir}

\begin{thebibliography}{99}

\bibitem[*]{byline} Present address: Physical and Theoretical 
Chemistry Laboratory, Oxford University, South Parks Road, Oxford OX1 3QZ. E-mail:
ederth@physchem.ox.ac.uk, fax: +44 (0)1865 275410, tel: +44 (0)1865 275400.

\bibitem{Casimir1948Attra}
H.~B.~G. Casimir, Proc. Kon. Ned. Akad. Wetensch., B {\bf 51},  793  (1948).

\bibitem{Mostepanenko1988Casim}
V.~M. Mostepanenko and N.~N. Trunov, Sov. Phys. Usp. {\bf 31},  965  (1988).

\bibitem{Milonni1994TheQu}
P.~W. Milonni, {\em The quantum vacuum} (Academic Press, Boston, 1994).

\bibitem{Lamoreaux1997Demon}
S.~K. Lamoreaux, Phys. Rev. Lett. {\bf 78},  5  (1997); {\bf 81}, 
5475 (1998).

\bibitem{Mohideen1998Preci}
U. Mohideen and A. Roy, Phys. Rev. Lett. {\bf 81},  4549  (1998).

\bibitem{Roy1999Impro}
A. Roy, C.-Y. Lin, and U. Mohideen, Phys. Rev. D {\bf 60},  111101  (1999).

\bibitem{Blocki1977Proxi}
J. Blocki, J. Randrup, W.~J. Swiatecki, and C.~F. Tsang, Ann. Phys. (New York)
  {\bf 105},  427  (1977).

\bibitem{White1983OnThe}
L.~R. White, J. Colloid Interface Sci. {\bf 95},  286  (1983).

\bibitem{Mehra1967Tempe}
J. Mehra, Physica {\bf 37},  145  (1967).

\bibitem{Schwinger1978Casim}
J. Schwinger, L.~L. DeRaad, and K.~A. Milton, Ann. Phys. (New York) {\bf 115},
  1  (1978).

\bibitem{Bezerra1997Casim}
V.~B. Bezerra, G.~L. Klimchitskaya, and C. Romero, Mod. Phys. Lett. {\bf 12},
  2613  (1997).

\bibitem{Klimchitskaya1999Compl}
G.~L. Klimchitskaya, A. Roy, U. Mohideen, and V.~M. Mostepanenko, Phys. Rev. A
  {\bf 60},  3487  (1999).

\bibitem{Lamoreaux1999Calcu}
S.~K. Lamoreaux, Phys. Rev. A {\bf 59},  3149  (1999).

\bibitem{Lifshitz1956Theor}
E.~M. Lifshitz, Soviet Physics JETP {\bf 2},  73  (1956), transl. from J.
  Exper. Theoret. Phys. USSR {\bf 29}, 94 (1955).

\bibitem{Klimchitskaya1996TheCor}
G.~L. Klimchitskaya and Y.~V. Pavlov, Int. J. Mod. Phys. A {\bf 11},  3723
  (1996).

\bibitem{Bordag1995TheCa}
M. Bordag, G.~L. Klimchitskaya, and V.~M. Mostepanenko, Int. J. Mod. Phys. A
  {\bf 10},  2661  (1995).

\bibitem{Mahanty1976Disper}
J. Mahanty and B.~W. Ninham, {\em Dispersion forces} (Academic Press, London,
  1976).

\bibitem{Parsegian1973vande}
V.~A. Parsegian and B.~W. Ninham, J. Theoret. Biol. {\bf 38},  101  (1973).

\bibitem{Palik1985Hand}
{\em Handbook of optical constants of solids}, edited by E.~D. Palik (Academic
  Press, Orlando, 1985).

\bibitem{Lambrecht1999Casim}
A. Lambrecht and S. Reynaud, Eur. Phys. J. D {\bf 8}, 309 (2000), quant-ph/9907105.

\bibitem{Israelachvili1985Inter}
J.~N. Israelachvili, {\em Intermolecular and Surface Forces}, 2nd ed. (Academic
  Press, London, 1985).

\bibitem{Wagner1995Forma}
P. Wagner, M. Hegner, H.-J. Guntherodt, and G. Semenza, Langmuir {\bf 11},
  3867  (1995).

\bibitem{Dubois1992Synth}
L.~H. Dubois and R.~G. Nuzzo, Annu. Rev. Phys. Chem. {\bf 43},  437  (1992).

\bibitem{Smith1980TheHy}
T. Smith, J. Colloid Interface Sci. {\bf 75},  51  (1980).

\bibitem{Atre1995Chain}
S.~V. Atre, B. Liedberg, and D.~L. Allara, Langmuir {\bf 11},  3882  (1995).

\bibitem{Parker1994Surfa}
J.~L. Parker, Prog. Surf. Sci. {\bf 47},  205  (1994).

\bibitem{Stewart1995TheUse}
A.~M. Stewart, Meas. Sci. Technol. {\bf 6},  114  (1995).

\bibitem{Johnson1971Surfa}
K.~L. Johnson, K. Kendall, and A.~D. Roberts, Proc. R. Soc. Lond. A {\bf 324},
  301  (1971).

\bibitem{Sridhar1997Adhes}
K.~L. Johnson, personal communication (the method is described in
I. Sridhar, K.~L. Johnson, and N.~A. Fleck, J. Phys. D: Appl. Phys. {\bf 30},
 1710  (1997)).

\bibitem{Thomas1999Water}
E. Thomas, Y. Rudich, S. Trakhtenberg, and R. Ussyshkin, J. Geophys. Res. {\bf
  104},  16053  (1999).

\bibitem{Tadros1974Adsor}
M.~E. Tadros, P. Hu, and A.~W. Adamson, J. Colloid Interface Sci. {\bf 49},
  184  (1974).

\bibitem{Awakuni1972Water}
Y. Awakuni and J.~H. Calderwood, J. Phys. D: Appl. Phys. {\bf 5},  1038
  (1972).

\bibitem{Parsegian1975LongR}
V.~A. Parsegian,  in {\em Physical chemistry: Enriching topics from colloid and
  surface science}, edited by H. van Olphen and K.~J. Mysels (Theorex, La
  Jolla, Calif., 1975), pp.\ 27--72.

\bibitem{Roth1996Impro}
C.~M. Roth and A.~M. Lenhoff, J. Colloid Interface Sci. {\bf 179},  637
  (1996).

\bibitem{Bostrom2000Comme}
M. Bostr\"om and B.~E. Sernelius, Phys. Rev. A {\bf 61},  046101  (2000).

\bibitem{Mohideen1999Reply}
U. Mohideen and A. Roy, Phys. Rev. Lett. {\bf 83},  3341  (1999).

\bibitem{Lamoreaux1999Comme}
S.~K. Lamoreaux, Phys. Rev. Lett. {\bf 83},  3340  (1999).

\bibitem{Derjaguin1975Effec}
B.~V. Derjaguin, V.~M. Muller, and Y.~P. Toporov, J. Colloid Interface Sci.
  {\bf 53},  314  (1975).

\bibitem{Greenwood1997Adhes}
J.~A. Greenwood, Proc. R. Soc. Lond. A {\bf 453},  1277  (1997).

\bibitem{Johnson1997AnAdh}
K.~L. Johnson and J.~A. Greenwood, J. Colloid Interface Sci. {\bf 192},  326
  (1997).

\end{thebibliography}
% \bibliographystyle{prsty}

% \bibitem{Sridhar1997Adhes}
% K.~L. Johnson, personal communication (the method is described in 
% I. Sridhar, K.~L. Johnson, and N.~A. Fleck, J. Phys. D: Appl. Phys. {\bf 30},
%   1710  (1997)).

\end{document}